\documentclass[a4paper]{article}

\usepackage{INTERSPEECH2020}
\usepackage{subfigure}
\usepackage{multirow}

\title{Power Pooling Operators and Confidence Learning for Semi-Supervised Sound Event Detection}
\name{Yuzhuo Liu$^{1,2}$,  Hangting Chen$^{1,2}$, Pengyuan Zhang$^{*1,2}$\thanks{
This work is partially supported by the National Natural Science Foundation of China (Nos. 11590774,11590772,11590770)
 }\thanks{*Pengyuan Zhang is the corresponding author.}
}

\address{
 $^1$Key Laboratory of Speech Acoustics \& Content Understanding, Institute of Acoustics, CAS, China\\
 $^2$University of Chinese Academy of Sciences, Beijing, China
}

\email{liuyuzhuo@hccl.ioa.ac.cn, chenhangting@hccl.ioa.ac.cn, zhangpengyuan@hccl.ioa.ac.cn}

\begin{document}

\maketitle
\begin{abstract}
In recent years, the involvement of synthetic strongly labeled data, weakly labeled data and unlabeled data has drawn much research attention in semi-supervised sound event detection (SSED). Self-training models carry out predictions without strong annotations and then take predictions with high probabilities as pseudo-labels for retraining. Such models have shown its effectiveness in SSED. However, probabilities are poorly calibrated confidence estimates, and samples with low probabilities are ignored. Hence, we introduce a method of learning confidence deliberately and retaining all data distinctly by applying confidence as weights. Additionally, linear pooling has been considered as a state-of-the-art aggregation function for SSED with weak labeling. In this paper, we propose a power pooling function whose coefficient can be trained automatically to achieve nonlinearity. A confidence-based semi-supervised sound event detection (C-SSED) framework is designed to combine confidence and power pooling. The experimental results demonstrate that confidence is proportional to the accuracy of the predictions. The power pooling function outperforms linear pooling at both error rate and $F_1$ results. In addition, the C-SSED framework achieves a relative error rate reduction of 34$\%$ in contrast to the baseline model.  
\end{abstract}

\section{Introduction}

Sound event detection (SED) is a task for identifying the categories and timestamps of target sound events in continuous audio recordings. As one of the core technologies in non-verbal sound perception and understanding, SED is widely deployed in various applications, such as noise monitoring for smart cities \cite{Bello2018SONYC}, nocturnally migrating bird detection \cite{8461410}, surveillance systems \cite{CroccoAudio} and multimedia indexing \cite{7952132}. It is time consuming to add high-quality labels to SED data manually. In comparison, synthetic strongly labeled data, weakly labeled data with clip-level categories only and unlabeled data are widely available. Therefore, research and competitions \cite{Serizel2018,Turpault2019} are turned to multiple instance learning (MIL) for SED with weak labeling and semi-supervised sound event detection (SSED) with the above data.

MIL \cite{AmoresMultiple} in SED permits models to learn frame-level classification from clip-level class labels. At each frame, a SED model predicts the probabilty of each sound event class being active. Then, a pooling function aggregates the frame-level predictions into a clip-level prediction for each sound event class. The derivative of the pooling function determines the direction of the frame-level gradient during backpropagation. The drawback of max pooling \cite{7952264} is that only one frame receives a non-zero gradient. The fact may cause many frame-level false negatives. The gradients of mean pooling \cite{RajA} and exponential softmax pooling \cite{expsoftmax} are always positive. Consequently, all frame-level predictions are boosted and many false positive results are produced for positive clips. The auto-pooling \cite{8434391} introduces a trainable parameter and can interpolate between the above three functions. A group of frame-wise trainable pooling weights are proposed in attention-based pooling \cite{8461975}. However, the gradients of auto-pooling and attention-based pooling are positive too. Linear pooling function \cite{linear} has been confirmed to work best for frame-level classification \cite{8682847}, since the gradient is positive where the frame-level probabilities are larger than half of the clip-level probability. However, ratio $1/2$ may not be the optimal value for various models and data. Hence, we propose an adaptive pooling function termed as power pooling, which can automatically learn the proper ratio.

SSED is to complete SED task with data partially annotated. Most SSED research is based on two classic semi-supervised learning methods: mean teacher and self-training. mean teacher \cite{TarvainenMean} averages model weights to form a target-generating teacher mode and achieves competitive results. A modified mean teacher model benefits SSED by employing both frame-level and clip-level consistency loss \cite{9000951,9053073,mt18,mt19}. Self-training \cite{selftraining}, a simple but effective bootstrapping semi-supervised method, cycles retraining the model with part of its own predictions as pseudo-labels. Self-training methods adopted in SSED \cite{USTC,Elizalde2017An,Joint} retrained only once and employed a small part unlabeled data with high probabilities. These approaches filtered unlabeled data by posterior to ensure the quality of the pseudo-labels, but caused three problems. First, Probability is not a calibrated indicator for evaluating the correctness of the model predictions. As modern neural network classifiers are designed to produce output probabilities prone to extreme values, incorrect predictions can be generated with high probabilities \cite{GuoOn}. Second, Simplified self-training methods   \cite{USTC,Elizalde2017An,Joint} lose considerable information. Third, These methods ignore true negative predictions. Nevertheless, data for SSED is extremely imbalanced. Massive correct negative predictions are beneficial to the retraining process. 

\begin{figure*}[t]
  \centering
  \includegraphics[width=\linewidth ]{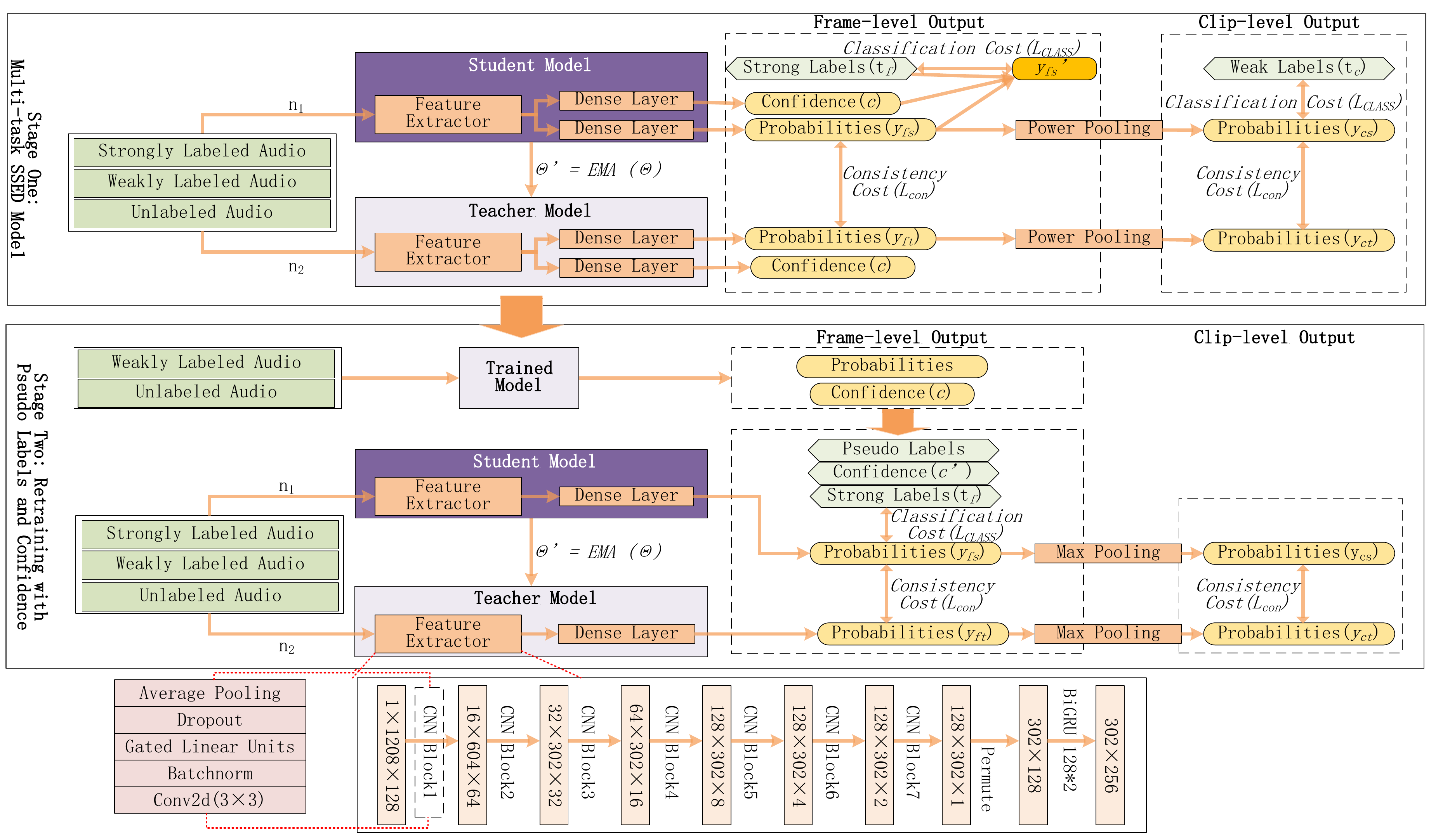}
  \caption{Framework C-SSED: Stage one introduces a muti-task system that can generate frame-level classification predictions and corresponding confidence estimates. The pseudo-labels, confidence for weakly labeled data, unlabeled data are produced and applied in stage two. The power pooling function is adopted in the first stage. $n_1$, $n_2$ are noises added to the student model and the teacher model. After the weights of the student model ($\Theta$) have been updated with gradient descent, the teacher model weights ($\Theta'$) are updated as an exponential moving average (EMA) of the student weights.}
  \label{fig:general}
\end{figure*}

Aiming at solving the above problems, this paper proposes a confidence-based semi-supervised sound event detection (C-SSED) framework (Figure \ref{fig:general}). C-SSED includes two stages of prediction and their corresponding confidence estimate generation and exploitation. Inspired by \cite{confidence}, C-SSED  simultaneously learns to detect sound events and assesses the correctness of predictions in the first stage. The system outputs frame-level classification predictions and confidence estimates for each sound event class. In the second stage, the predictions produced in stage one are used as pseudo-labels for weakly labeled and unlabeled data. The confidence is used to weight the frame-level classification loss. All data are distinctly retrained once. Since mean teacher has shown its strength in the SSED task, we adopted it in C-SSED as the backbone model. The contributions of this paper include: 

1. We develop a power pooling function that is simple to implement, and requires little additional computation over the state-of-the-art linear pooling. Additionally, it outperforms linear pooling at both error rate (ER) and $F_1$. 

2. We design a C-SSED framework. Confidence, a novel prediction measurement and its usage are introduced in the C-SSED framework. Our empirical results show that confidence can reflect the accuracy of predictions. Moreover, our usage realizes retraining all data using pseudo-labels without cycling training and effectively decreases the ER.
 
3. The C-SSED framework successfully adds self-training to mean teacher method. The ER decreases compared to adapting mean teacher only.

\section{Proposed System}
\subsection{Baseline}

Mean teacher is a consistency regularization method that evaluates unlabeled data with two different noises, and then apply a consistency cost between the two predictions. In this case, the model assumes a dual role as a \emph{teacher} and a \emph{student}. The baseline model is performed the following optimization on the basis of mean teacher model in \cite{mt18}. First, we introduce data augmentation by shifting input features along the time axes (forward and backward with a Normal distribution with zero mean and a standard deviation of 16 frames). Second, we adopt a set of median filter window sizes that is proportional to the average duration of different event categories \cite{9053584}. Third, the 128-dimensional log mel-spectrogram is extracted at each frame. The size of the window is 2048 and the hop length is 365. Fourth, parameters of feature extractor follow the settings in \cite{mt19}.

\subsection{Pooling Function}
\subsubsection{Linear Pooling}

\begin{table}[t]
  \caption{ The clip-level and frame-level gradient direction of linear pooling. $y_f$ and $y_c$ are frame-level and clip-level predictions. }
  \label{tab:linear}
  \centering
  \resizebox{0.45\textwidth}{!}{
  \begin{tabular}{ccccccc}
    \toprule
    \textbf{Label}         & \textbf{Clip-level}        & \textbf{Condition}             & \textbf{Frame-level}        \\         
    \hline                                             
     positive ($t=1$)             & $y_c\rightarrow1$    &$ y_f>y_c/2$          & $y_f\rightarrow1$  \\
                                  &                      &$ y_f < y_c/2$        & $y_f\rightarrow0$ \\
    \hline 
     negative ($t=0$)            & $y_c\rightarrow0$     &$ y_f>y_c/2$          & $y_f\rightarrow y_c/2$  \\
                                 &                       &$ y_f < y_c/2$        &  $y_f\rightarrow y_c/2$ \\

    \bottomrule
  \end{tabular}}
  
\end{table}

The formulas of linear pooling function and its gradient can be written as:
\begin{equation}\label{equ:linear}
  y_c = \frac{\sum\nolimits_{i}y_f(i) \times y_f(i)}{\sum\nolimits_{i}y_f(i)},
\end{equation}
 
\begin{equation}\label{equ:linear gradient}
  \frac{\partial y_c}{\partial y_f(i)} = \frac{2y_f(i) - y_c}{\sum\nolimits_{j}y_f(j)}.
\end{equation} 
The clip-level and frame-level gradient directions are illustrated in Table \ref{tab:linear}. As a result, when $t=1$, larger $y_f$ is driven to 1 and smaller $y_f$ is driven to 0, benefiting the timestamps detection. When $t=0$, $y_f$ is pushed towards $y_c/2$. Considering that $y_c$ is a weighted average of the $y_f$, all the $y_f$ will converge to 0 as desired after enough iterations.

\subsubsection{Power Pooling}

For positive recordings ($t=1$), linear pooling function leads larger $y_f$ to be boosted. Limited by the form of linear function, the threshold of larger $y_f$ is defined as $y_c / 2$. However, the threshold is supposed to be adjusted dynamically according to the value of $y_c$ and the number of positive frame-level samples. This issue can be addressed by applying a power pooling function:
\begin{equation}\label{equ:power}
  y_c = \frac{\sum\nolimits_{i}y_f(i) \times y_f^{n}(i)}{\sum\nolimits_{i}y_f^{n}(i)},
\end{equation}
and the gradient can be written as:
\begin{equation}\label{equ:power gradient}
  \frac{\partial y_c}{\partial y_f(i)} = \frac{(n+1) \times y_f^{n}(i)-n\times y_f^{n-1}(i) \times y_c}{\sum\nolimits_{j}y_f^{n}(j)},
\end{equation}  
parameter  $n\neq-1$ and threshold $\theta = n/(n+1)$ ($\theta$ $\neq1$ ). Treating $n$ as a free parameter to be learned along-side the model parameters allows eq.(\ref{equ:power}) to automatically adapt to and interpolate between separate pooling functions. For instance, when $n=0$, eq.(\ref{equ:power}) reduces to mean pooling. When $n=1$, eq.(\ref{equ:power}) simplifies to linear pooling. When $n\rightarrow\infty$, eq.(\ref{equ:power}) approaches the max aggregation.

We discuss the diversification of $y_c$ and $y_f$ more specifically in three cases according to the value of $n$. For $n\in(0,+\infty)$, $\theta\in(0,1)$, weight $y_f^{n}$ increases where $y_f$ increases, leading to $y_c$ being generated under the \emph{standard multiple instance} (SMI) \emph{assumption}: the bag label is positive if and only if the bag contains at least one positive instance. During backpropagation,  parameters for producing frame-level outputs upgrade as linear pooling pattern (Table\ref{tab:linear}) with dynamic threshold $\theta$ instead of value $1/2$. For $n\in(-1,0]$, $\theta\in(-\infty,0]$, $(\theta \times y_c)\leq0$. Therefore, for negative clips ($t=0$), all $y_f$ are pushed towards 0, as desired. Nevertheless, for positive clips ($t=1$), all $y_f$ increase towards 1 as the mean pooling mode. For $n\in(-\infty,-1)$, $\theta\in(1,+\infty)$, weight $y_f^{n}$ increases where $y_f$ decreases. This behaviour violates the SMI assumption. A positive $y_c$ can only be produced when the vast majority $y_f$ is positive. 

\subsection{C-SSED}
\subsubsection{Stage One: Multi-task SSED Model (MT-SSED)}

To get reliable confidence, we added a branch to train confidence in MT-SSED. When solving the issue of simultaneously generating SED predictions and their corresponding confidence without the confidence label, we draw on the successful experience in the field of out-of-distribution detection \cite{confidence}. The motivation is equivalent to a special test that permits giving hints. Candidates are allowed to ask for hints according to their confidence of the questions. Furthermore, a certain penalty is carried out in order to prevent candidates from tending to ask for hints for all questions. For obtaining the highest score, candidates must improve their ability to answer questions and self-assess at the same time. 

MT-SSED is constructed based on the baseline model. There are four outputs in the baseline model, the frame-level output $y_{ft}$ and clip-level output  $y_{ct}$ of the teacher model, the frame-level output $y_{fs}$ and clip-level output $y_{cs}$ of the student model. We choose power pooling as the aggregation function in MT-SSED. To make the model self-assess, we add a confidence branch in parallel with the original
class prediction branch. The confidence branch, which shares the same structure with the frame-level classification branch, applies a fully-connected layer followed by sigmoid. The confidence branch generates corresponding confidence values $c$ for the classification results of each sound event at every frame. Output $c$ takes values between 0 and 1. If the model is confident about the classification, output $c$ will be closer to 1. Conversely, if the model is uncertain about the correctness of classification predictions, the value of $c$ will be closer to 0.

A crucial issue of confidence is how to achieve the training of two tasks with just the classification labels. Following the main idea of giving hints, we construct a new frame-level output of student model $y_{fs}'$ with the label $t_f$ and two outputs $y_{fs}$ and $c$:
\begin{equation}
  y_{fs}' = (1-c) \times t_f+ c \times y_{fs}.
\end{equation}
The outputs of the student model $y_{fs}'$ and $y_{cs}$ are in comparison with strong labels $t_f$ and $t_c$ utilizing the binary cross entropy (BCE) loss. The classification loss can be written as 
\begin{equation}
\begin{aligned}
L_{class} &=L_{class_f}+L_{class_c}{} \\
	      & = L_{BCE}(y_{fs}', t_f)+L_{BCE}(y_{cs},t_c).
\end{aligned}	   
\end{equation}	   
Outputs $y_{fs}$ and $y_{cs}$ are compared with the outputs $y_{ft}$ and $y_{ct}$ by applying the mean square error (MSE) loss. The consistency loss is
\begin{equation}
    L_{con} = L_{MSE} (y_{ft},y_{fs}) + L_{MSE}(y_{ct},y_{cs}).
\end{equation}
	
Training with $L_{class}$ and $L_{con}$ loss functions, the network will be lazy to learn the differences between classes. Instead, the model tends to make $c$ approach 0 and receives ground truth for every sample. Thus, a log penalty is added to the loss function. The confidence loss can be interpreted as a BCE loss:
\begin{equation}
  L_c = -log(c).
\end{equation}

The loss function of the multi-task system is
\begin{equation}
  L = L_{class} + \mu \times L_{con} + \lambda \times L_c,
\end{equation}
parameter $\mu$ increases with epochs and $\lambda$ is a hyperparameter. When $\lambda$ is too small, MT-SSED model tends to ask for hints and performs poorly in classification. When $\lambda$ is too large, the confidence $c$ $\rightarrow1$ and lose the distinction. To ensure the effects of both classification and confidence estimation, we first optimize the mean teacher model and classification branch without $L_c$. Then, the trained parameters are fixed, and $L$ is deployed to train the confidence branch separately for 5 epochs. 

\subsubsection{Stage Two: Retraining with Pseudo-Labels and Confidence}

In the second stage, the weakly labeled and unlabeled data are sent to the trained MT-SSED model to yield frame-level predictions and confidence estimates. The frame-level posterior probabilities are applied as soft pseudo-labels for the above data during retraining. Confidence estimates offer a model's self-assessment for pseudo-labels. For weakly labeled data, we regulate the outputs. Pseudo-labels are revised to 0, and confidence estimates are revised to 1 for negative clips. For strongly labeled data, all confidence estimates are set to 1. In order to guarantee the contribution of each sample under the premise of discrimination, we interpolate the confidence values with a hyperparameter $\alpha$ to produce a new confidence:
\begin{equation}
  c' = \alpha \times c+(1-\alpha) \times 1.
\end{equation}

The frame-level classification loss $L_{class_f}$ is weighted by $c'$ as follows:
\begin{equation}
	L_{class_f} = \frac{\sum\nolimits_{i,k}c'(i,k) \times L_{BCE}(y_{fs}(i,k),t_f(i,k)))}{\sum\nolimits_{i,k}c'(i,k)},
\end{equation}
where $i,k$ represents the index of frames and classes. For those samples with high confidence, the accuracy of their pseudo-labels is higher. We make them more important during retraining. Conversely, the proportion of classification loss function value is relatively small for other samples. As a result, all strongly labeled, weakly labeled and unlabeled data information are learned distinctly. $L_{class_c}$ and $L_c$ are omitted. Since there is no weak label to guide the training of frame-level predictions in stage two, the max pooling is adopted to produce clip-level predictions. 

\section{Experiments and Discussion}
\subsection{Dataset and Metrics}

We carried our experiments on the DCASE2019 task4 dataset \cite{Turpault2019}. The dataset can be divided into four subsets, including training sets (synthetic strongly labeled: 2,045 clips, weakly labeled: 1,578 clips, unlabeled: 14,412 clips) and validation set (1,168 clips). In addition, the evaluation set of DCASE2018 (800 clips) \cite{Serizel2018} was tested in the C-SSED experiment. The duration of each audio clip is 10 seconds, and multiple audio events may occur at the same time. The sampling rate is 44100 Hz.

Experiments were evaluated mainly with event-based macro-average error rate (ER) with a 200 ms collar on onsets and a 200 ms/20$\%$ of the events length collar on offsets. Both event-based and segment-based (1 s) ER and $F_1$ were applied in the pooling function experiments. A smaller ER is better and a larger $F_1$ is better. The specific evaluation details can be found in \cite{Mesaros2016_MDPI}.

\subsection{Results and Analysis}
\subsubsection{Pooling Function}
 \begin{figure}[t]
  \centering 
  \resizebox{0.3\textwidth}{!}{
  \includegraphics[width=\linewidth]{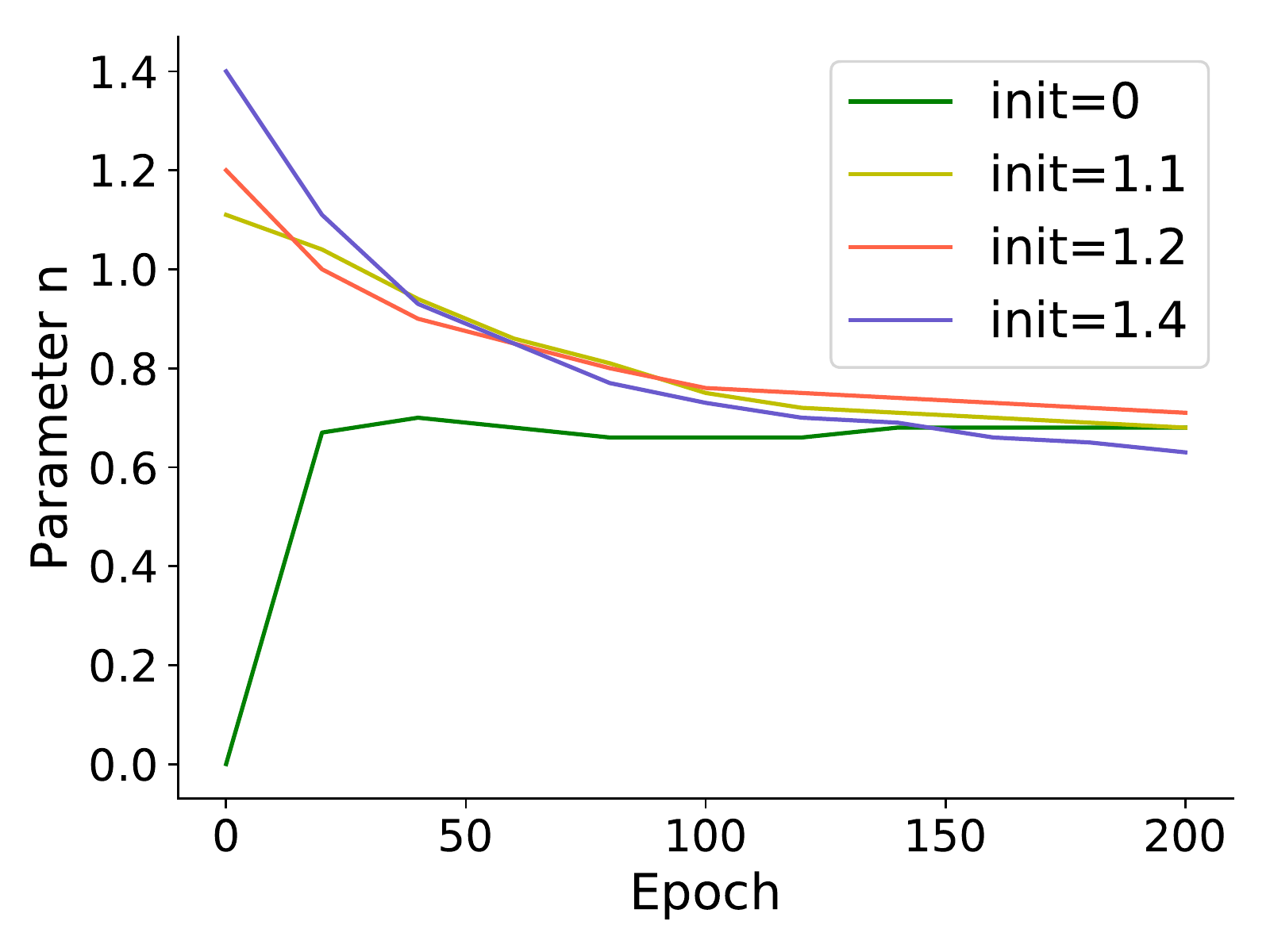}}
  \caption{The parameter $n$ of power pooling in different epochs with different initial values (init).}
  \label{fig:power_init}
\end{figure}

We observe that the initialization of the parameter $n$ influenced the training process. If the value of $n$ was too large (for instance, the initial value was 10), only very few frame-level samples could be updated towards positive. The change for clip-level predictions was slow as they were generated from frame-level predictions. The training was trapped at a low level. Therefore, the initial value of $n$ is recommended to be between [0, 3]. 

\begin{table}[t]
\caption{Detailed performance of four pooling functions. Initial value of power pooling is 1.2.}
  \label{tab:power}
  \centering
 \resizebox{0.48\textwidth}{!}{
 \begin{tabular}{r|l|ll|ll}
\toprule
& & \multicolumn{2}{l|}{\textbf{Event-Based}} & \multicolumn{2}{l}{\textbf{Segment-Based}} \\ \cline{3-6} 
& \textbf{Pooling Function}                                                                                  & \textbf{ER}($\%$)        & $\bm{F_1}$($\bm{\%}$)        & \textbf{ER}($\%$)         & $\bm{F_1}$($\bm{\%}$)          \\ \hline
\textbf{Attention}          &$y_c=\frac{\sum\nolimits_{i}y_f(i)\times \omega_i}{\sum\nolimits_{i}\omega_i}$                                            & 1.26               & 32.04               & 0.72                & 35.75                \\ \hline
\textbf{Auto}  &$y_c=\frac{\sum\nolimits_{i}y_f(i)\times \exp(\beta\times y_f(i))}{\sum\nolimits_{i}\exp(\beta\times y_f(i))}$  & 1.16               & 26.15               & 0.67                & 63.14                \\ \hline
\textbf{Linear}             &$y_c=\frac{\sum\nolimits_{i}y_f(i)\times y_f(i)}{\sum\nolimits_{i} y_f(i)}$  & 1.08               & 34.27               & 0.68                & 62.32                \\ \hline
\textbf{Power}              &$y_c=\frac{\sum\nolimits_{i}y_f(i)\times y_f^{n}(i)}{\sum\nolimits_{i} y_f^{n}(i)}$  & \textbf{1.07}      & \textbf{37.04}      & \textbf{0.64}       & \textbf{63.57}  \\ 
\bottomrule   
 
\end{tabular}}
\end{table}
                                      
Figure \ref{fig:power_init} illustrates that parameter $n$ eventually stabilized at approximately 0.7 regardless of the initial value. This phenomenon proves that the power pooling can adaptively learn a certain value with the given model architecture and dataset. As listed in Table \ref{tab:power}, the power pooling outperformed attention pooling \cite{mt19}, auto-pooling \cite{8434391} and linear pooling \cite{linear}. Power pooling achieved an event-based $F_1$ score equal to $37.04\%$, exhibiting a relative improvement of $15.6\%$, $41.64\%$ and $8.08\%$ compared with baseline methods. The results hint that the certain value of parameter $n$ learned by the power pooling is more proper than $n=1$ in linear pooling. Furthermore, the results of linear pooling are slightly worse than auto-pooling on segment-based indicators. The power pooling is well suitable for both event-based and segment-based SSED. 

\subsubsection{C-SSED} 
\begin{figure}[t]
  \centering
    \subfigure[]{
	\begin{minipage}[t]{0.48\linewidth}
	  \setlength{\abovecaptionskip}{-2cm} 
	\centering 
	\includegraphics[width=\linewidth]{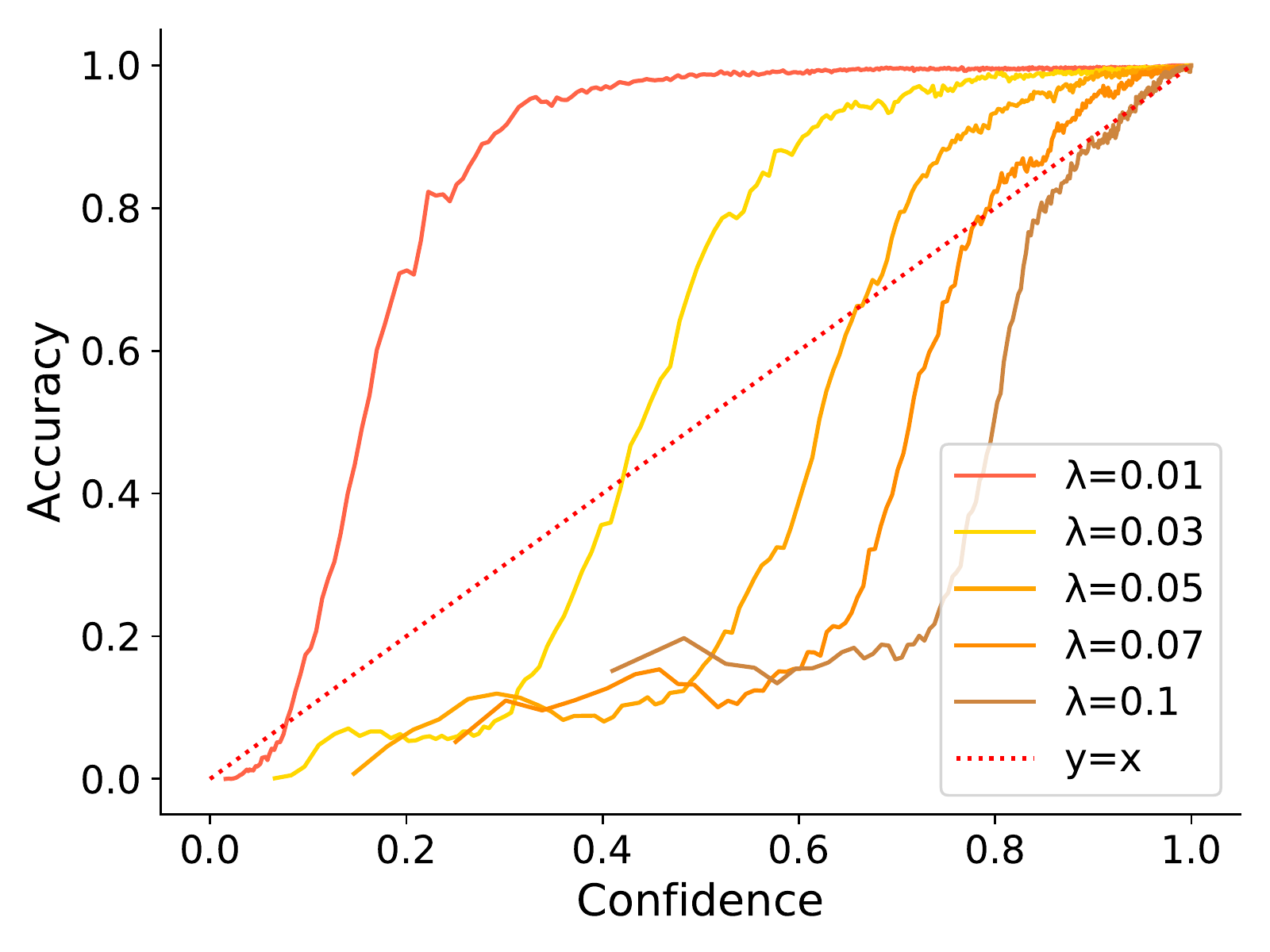}
	\label{fig:lambda}
	\end{minipage}%
	}%
  	\subfigure[]{
	\begin{minipage}[t]{0.48\linewidth}
	\centering
	\includegraphics[width=\linewidth]{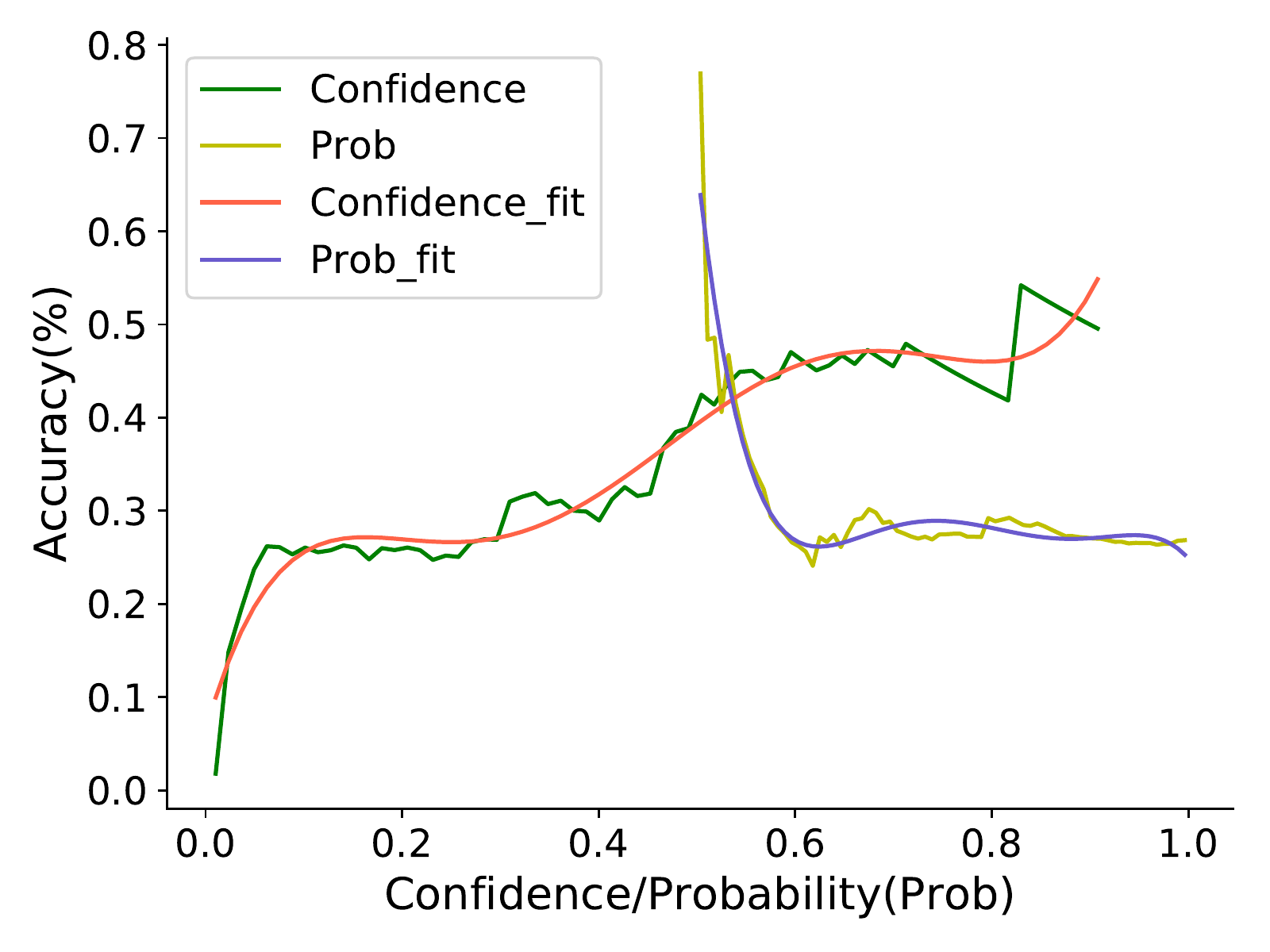}
	\label{fig:prob}
	\end{minipage}%
	}%
\centering
\caption{ Accuracy of frame-level predictions changes with confidence estimates and probabilities. (a) The accuracy of all labels was obtained from models trained using different values of parameter $\lambda$. (b) The accuracy of positive predictions ($prob\geq0.5$) changed with learned confidence estimations and probabilities. Their polynomial fitting curves are displayed. Here, the DCASE2019 validation set was tested by the MT-SSED model.}
\end{figure}

As described in Section 2.3.1, parameter $\lambda$ greatly influences the threshold and distinctiveness of confidence. We adjusted $\lambda$ in a small range. Figure \ref{fig:lambda} illustrates that all MT-SED models trained with $\lambda\in[0.01,0.1]$ could generate confidence that is positively related to the accuracy of predictions. However, if $\lambda$ is relatively large, the confidence values are aggregated. If $\lambda$ is relatively small, majority confidence estimates possess the similar accuracy. Polyline $y=x$ represents the ideal discrimination of prediction quality measurement. In our experiment, curve $\lambda=0.03$ brought out the best distinguishment.

Figure \ref{fig:prob} focuses on positive predictions, as classic self-training methods choose samples with high probabilities. Figure \ref{fig:prob} demonstrates that accuracy increases with confidence estimates. For posterior probabilities, accuracy first decreases rapidly and then flattens. The reason is that the classification outputs are trained towards 0 or 1. The number of samples with probabilities in [0.5,0.6] is relatively small, which might result in high accuracy. Most posterior probabilities are concentrated in [0.9,1], but they contain many false positive frames. Figure \ref{fig:prob} confirms that the posterior probabilities cannot evaluate the correctness of predictions. In contrast, we notice that the confidence of positive predictions is concentrated in [0,0.1]. This behaviour coincides with massive false positives produced by MT-SSED. Figure \ref{fig:prob} and Figure \ref{fig:lambda} indicate that the confidence can measure the accuracy of predictions.

\begin{table}[t]
  \caption{Comparison of models in terms of ER (in $\%$). Retraining with $\alpha=0$ is equal to retrain without confidence.}
  \label{tab:C-SSED}
  \centering
  \resizebox{0.45\textwidth}{!}{
\begin{tabular}{cccccccc}
\toprule
                                                &                          & \multicolumn{3}{c}{\textbf{Evaluation 2018}}  & \multicolumn{3}{c}{\textbf{Validation 2019}}   \\ \cline{3-8} 
\textbf{Model}                                  &                          & \textbf{ER} & \textbf{DEL} & \textbf{INS} & \textbf{ER} & \textbf{DEL} & \textbf{INS} \\ \hline
MT18                                 & \multicolumn{1}{c}{}    & 1.65        & 0.78         & 0.87         & 1.56        & 0.76         & 0.80         \\
Baseline                                         & \multicolumn{1}{c}{}    & 1.34        & 0.72         & 0.62         & 1.26        & 0.70          & 0.56         \\
MT-SSED                                           & \multicolumn{1}{c}{}    & 1.13        & 0.69         & 0.44         & 1.07        & 0.67         & 0.40          \\
\midrule
\multicolumn{1}{c}{\textbf{Retriain}}          & \multicolumn{1}{c}{$\bm{\alpha}$}  & \multicolumn{3}{c}{}                     & \multicolumn{3}{c}{}                     \\ \hline
                   & \multicolumn{1}{c}{0}   & \textbf{1.10}         & \textbf{0.68}         & \textbf{0.42}         & \textbf{1.04}        & \textbf{0.67}         & \textbf{0.37}         \\ \hline
\multicolumn{1}{c}{Prob0.9}                       & \multicolumn{1}{c}{1} &3.72          &0.70    &3.02         &3.41        &0.69          &2.72   \\
\multicolumn{1}{c}{Prob}                       & \multicolumn{1}{c}{0.3} & 1.15        & \textbf{0.68}         & 0.47         & 1.09        & \textbf{0.66}         & 0.43         \\  
\multicolumn{1}{c}{Prob0.5}                       & \multicolumn{1}{c}{0} &1.19         &\textbf{0.68}    &0.51      &1.14  &0.68  
&0.46 \\
\hline
\multicolumn{1}{c}{C-SSED}                     & \multicolumn{1}{c}{0.3} & \textbf{1.09}        & 0.70          &\textbf{0.39}         & \textbf{1.03}        & \textbf{0.68}         & \textbf{0.35}         \\
\multicolumn{1}{c}{($\lambda$=0.1)}                    & \multicolumn{1}{c}{0.7} & \textbf{1.13}        & \textbf{0.69}         & \textbf{0.44}         & \textbf{1.05}        & \textbf{0.67}         & \textbf{0.38}         \\
\multicolumn{1}{c}{}                           & \multicolumn{1}{c}{1}   & \textbf{1.06}        & \textbf{0.68}         & \textbf{0.38}         & \textbf{1.02}        & \textbf{0.67}         & \textbf{0.35}         \\ \hline
\multicolumn{1}{c}{C-SSED}                   & \multicolumn{1}{c}{0.3} & \textbf{1.08}        & \textbf{0.69}         & \textbf{0.39}         & \textbf{1.01}        & \textbf{0.67}         & \textbf{0.34}        \\
\multicolumn{1}{c}{($\lambda$=0.01)}                           & \multicolumn{1}{c}{0.7} & \textbf{1.06}        & 0.70         & \textbf{0.36}         & \textbf{1.01}        & 0.69         & \textbf{0.32}         \\
\multicolumn{1}{c}{}                           & \multicolumn{1}{c}{1}   & \textbf{1.00}  & 0.86         & \textbf{0.14} & \textbf{0.98}        & 0.85         & \textbf{0.13}       \\  
  \bottomrule
\end{tabular}}
\end{table}

We trained C-SSED models with different values of $\lambda$ and $\alpha$, and compared proposed models with the following approaches: \\
\emph{MT18}: the official baseline for DCASE2019 task4, with mean teacher structure \cite{mt18}.\\
\emph{Baseline}: modified MT18 method as described in 2.1 with attention pooling. \\
\emph{MT-SSED}: the stage one model of C-SSED with power pooling.\\
\emph{Prob0.9}: only predictions with $prob\geq0.9$ added to pseudo-labels, samples retrained with equal weight \cite{USTC}.\\
\emph{Prob}: all samples retrained with probabilities as weights.  \\
\emph{Prob0.5}: only predictions with $prob\geq0.5$ added to pseudo-labels, samples retrained with confidence.

Table \ref{tab:C-SSED} illustrates that C-SSED models are significantly improved in contrast to the other models. The ER improvement is mainly due to the significant reduction of INS error. That is, the improvement comes from correcting false positives. When the model retrained with parameter $\alpha=0$, ER also decreased. Although the mean teacher method already makes use of weakly labeled and unlabeled data, applying an appropriate self-training structure can effectively reduce false alarms. However, \emph{Prob0.9} introduced massive false positive predictions as pseudo-labels and applied them equally, resulting in many insertion errors. \emph{Prob} caused performance degradation due to the fact that  the majority true negatives owned small weights approximately 0.7 after applying parameter $\alpha=0.3$. Meanwhile, many false positives (Figure \ref{fig:prob}) were introduced with high weights. In contrast, C-SSED with confidence further improved ER by successfully strengthening the attention to true negatives and true positives. The poor results of \emph{Prob0.5} reveal the importance of true negatives. 

Table \ref{tab:C-SSED} and Figure \ref{fig:lambda} demonstrate that a suitable combination of parameters $\lambda$ and $\alpha$ can produce more balanced detection results. INS errors were reduced under the premise of small fluctuations in DEL errors. As an example, when $\lambda=0.1$, $\alpha=1$ or $\lambda=0.01$, $\alpha=0.7$, the total number of events predicted correctly did not reduce while ER decreased significantly. These combinations increase the lower bound of confidence values and ensure the information contribution of samples with low confidence. Meantime, preserve the distinction of confidence. 

\section{Conclusions}

The power pooling function proposed in this paper benefits SSED with weak labeling by maintaining the optimal frame-level parameter updating scheme and adaptively learning the suitable threshold. Moreover, it can smoothly interpolate between various pooling functions, such as max, mean, and linear pooling. While this paper focuses on SSED applications, power pooling is general, and can be utilized widely to MIL problems in any domain. 

Using the power pooling function, we build a C-SSED framework. Our experiments verified that the proper combination of self-training and mean teacher method is better than employing mean teacher alone. The confidence estimates are used as weights to optimize the self-training retraining process, which creates a further improvement. The C-SSED framework can be extended to other semi-supervised tasks. In addition, this paper introduces a confidence training method to SSED, but confidence can also be applied in other scenes, such as optimizing focal loss.

\bibliographystyle{IEEEtran}

\bibliography{mybib}

\end{document}